\documentclass[rnote]{aa}
\usepackage{graphicx}
\usepackage{natbib}

\usepackage{color}

\usepackage{amssymb}

\usepackage[varg]{txfonts}

\begin{document}

\definecolor{bleu}{rgb}{.3,0.2,1.0}
\definecolor{red}{rgb}{1.,0.2,0.2}

\title{A distinct magnetic property of the inner penumbral boundary}
\subtitle{II. Formation of a penumbra at the expense of a pore}

\author{J. Jur\v{c}\'{a}k
        \inst{1}
        \and
        N. Bello Gonz\'{a}lez
        \inst{2}
        \and
        R. Schlichenmaier
        \inst{2}
        \and
        R. Rezaei
        \inst{2, 3, 4}}

\institute{Astronomical Institute of the Academy of Sciences, Fri\v{c}ova  298, 25165 Ond\v{r}ejov, Czech Republic
  \and
  Kiepenheuer-Institut f\"{u}r Sonnenphysik, Sch\"{o}neckstr. 6, 79104 Freiburg, Germany
  \and
  Instituto de Astrof\'isica de Canarias (IAC), V\'ia Lact\'ea, 38200 La Laguna (Tenerife), Spain 
  \and
  Departamento de Astrof\'isica, Universidad de La Laguna, 38205 La Laguna (Tenerife), Spain}

\date{Received 11 December, 2014; accepted }

\abstract
{We recently presented evidence that stable umbra-penumbra boundaries are characterised by a distinct canonical value of the vertical component of the magnetic field, $B^{\rm stable}_{\rm ver}$. In order to trigger the formation of a penumbra, large inclinations in the magnetic field are necessary. In sunspots, the penumbra develops and establishes by colonising both umbral areas and granulation, that is, penumbral magneto-convection takes over in umbral regions with $B_{\rm ver} < B^{\rm stable}_{\rm ver}$, as well as in granular convective areas. Eventually, a stable umbra-penumbra boundary settles at $B^{\rm stable}_{\rm ver}$.}
  {Here, we aim to study the development of a penumbra initiated at the boundary of a pore, where the penumbra colonises the entire pore ultimately.}
 {We have used Hinode/SOT G-band images to study the evolution of the penumbra. Hinode/SOT spectropolarimetric data were used to infer the magnetic field properties in the studied region.}
{The penumbra forms at the boundary of a pore located close to the polarity inversion line of NOAA\,10960. As the penumbral bright grains protrude into the pore, the magnetic flux in the forming penumbra increases at the expense of the pore magnetic flux. Consequently, the pore disappears completely giving rise to an orphan penumbra. At all times, the vertical component of the magnetic field in the pore is smaller than $B^{\rm stable}_{\rm ver} \approx 1.8$~kG.}
{Our findings are in an agreement with the need of $B^{\rm stable}_{\rm ver}$ for establishing a stable umbra-penumbra boundary: while $B_{\rm ver}$ in the pore is smaller than $B^{\rm stable}_{\rm ver}$, the protrusion of penumbral grains into the pore area is not blocked, a stable pore-penumbra boundary does not establish, and the pore is fully overtaken by the penumbral magneto-convective mode. This scenario could also be one of the mechanisms giving rise to orphan penumbrae.}
  
\keywords{ Sun: magnetic fields --
           Sun: photosphere --
           Sun: sunspots
               }

\maketitle

%
%

\section{Introduction}
\label{introduction}

Penumbra formation is not yet fully understood. \citet{Rucklidge:1995} showed that the efficiency of energy transport across the magnetopause increases dramatically when the magnetic field inclination exceeds the critical value of 45$^\circ$ ($\gamma_{\rm crit}$). Despite the simplicity of the model used, the $\gamma_{\rm crit}$ of 45$^\circ$ for the penumbral formation was confirmed by recent MHD simulations of \citet{Rempel:2009b}. \citet{jurcak:2014a} confirmed observationally the importance of magnetic field inclination in triggering penumbra formation. 

The magnetic field inclination on a protospot boundary is related to the total magnetic flux as the magnetic field inclination increases with the increasing flux. According to theoretical prediction of \citet{Rucklidge:1995}, the necessary magnetic flux for a penumbra formation is between $1-7 \times 10^{20}$~Mx. These values are in agreement with observed values of $5 \times 10^{20}$~Mx \citep{Zwaan:1987}, $1 \times 10^{20}$~Mx \citep[rudimentary penumbra,][]{Leka:1998}, and $4 \times 10^{20}$~Mx \citep[protospot with forming penumbral segments,][]{Rezaei:2012}.

Observations of orphan penumbrae \citep{Zirin:1991, Kuckein:2012, Lim:2013, jurcak:2014, Zuccarello:2014} show that they are comparable in all aspects to regular sunspot penumbrae. These observations prove that sufficient magnetic flux is not a necessary condition for a penumbra formation. A favourable configuration of magnetic field strength and inclination can result in the penumbra formation. 

Once the penumbra formation is triggered, the penumbra extends mostly at the expense of granular regions. \citet{Jurcak:2015} show that the inner penumbral end extends also to the umbral area where the umbra-penumbra (UP) boundary settles at a distinct canonical value of the vertical component of the magnetic field, $B^{\rm stable}_{\rm ver}$ of 1.8~kG. This confirms the results of \citet{Jurcak:2011} who found constant values of $B_{\rm ver}$ along UP boundaries of stable sunspots, where the actual value of $B^{\rm stable}_{\rm ver}$ seems to be weakly dependent on the sunspot size. Larger samples of sunspots must be investigated to clarify whether or not such a dependence is real.

\begin{figure*}[!t]
 \centering \includegraphics[width=0.95\linewidth]{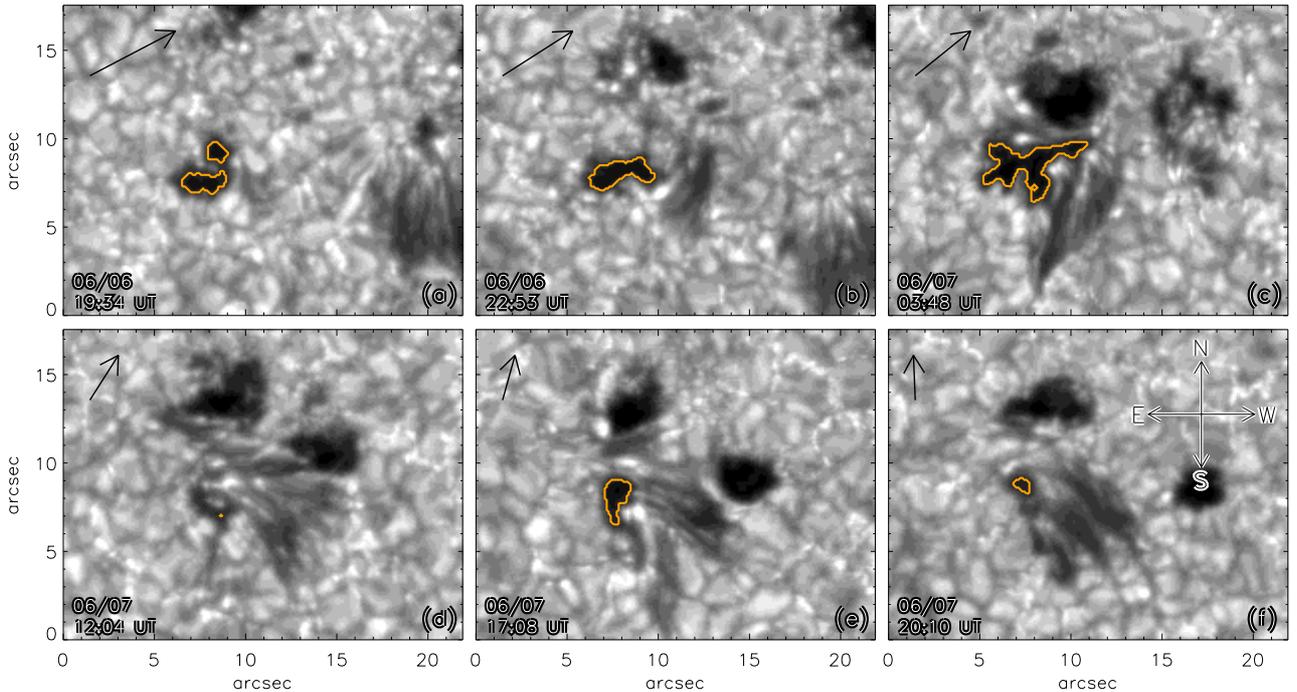}
 \caption{G-band images showing selected stages of pore and penumbra evolution in AR~10960. The orange contours indicate pore areas with $I_\mathrm{c} < 0.5 \times I^\mathrm{QS}_\mathrm{c}$. The arrows point towards the disc centre (north is up and west is right as marked in f). A video showing the temporal evolution of the penumbra at the pore boundary is available in the online edition.}
 \label{G-band}
\end{figure*}

In this paper, we present the Hinode observations of a small pore near the polarity inversion line of AR~10960. We follow the formation and development of a penumbra forming at the expense of the magnetic flux of the pore while colonising it. Six spectro-polarimetric scans of the region are used to infer the maps of the magnetic field vector. Details of the studied data are described in Sect.~\ref{observations} along with the analysis methods. We describe and discuss the results in Sect.~\ref{results}, and summarise them in Sect.~\ref{discussion}. 

\section{Observations and data analysis}
\label{observations}

Our analysis is based on data observed with the Solar Optical Telescope \citep[SOT,][]{Tsuneta:2008} aboard the Hinode satellite \citep{Kosugi:2007}. We made use of spectro-polarimetric (SP) data in the \ion{Fe}{I} lines at 630.15 and 630.25~nm, and images taken through a G-band filter centred at 430.5~nm with a bandpass of 0.8~nm. 

We study the evolution of a pore in AR~10960 between 19:00~UT on 6 June 2007 and 8:00~UT on 8 June 2007. The pore was located 6$^\circ$S and between 14$^\circ$E and 6$^\circ$W during the analysed time period. The cadence of G-band images vary between 60~sec and 100~sec (see Fig.~\ref{pore_area}). At all times, the spatial sampling is 0\farcs11 resulting in a diffraction-limited spatial resolution of 0\farcs22. During the analysed time period, six Hinode/SP scans of the region were acquired. The SP observations were taken in a fast mode with a pixel sampling of 0\farcs32 and a noise level of $10^{-3}I_{\rm c}$. These data were calibrated using the available routines in the Hinode SolarSoft package.

In order to retrieve the physical parameters, the SP data were inverted with the SIR code \citep[Stokes Inversion based on Response function,][]{Cobo:1992}. We used a simple atmosphere model supposing all free parameters of the inversion, except for temperature, to be constant with height. We took into account the spectral point spread function of the Hinode SP in the inversion process. We assumed the magnetic filling factor to be unity and assumed no stray light. The macro-turbulence was set to zero while micro-turbulence was a free parameter of the inversion. We applied the code AMBIG \citep{Leka:2009} to solve the 180$^\circ$ ambiguity of the LOS azimuth. The disambiguated LOS vector magnetic field was then transformed to the local reference frame (LRF) with the help of routines from the AZAM code \citep{Lites:1995}.

The apparent motions of photospheric structures were determined from the G-band images using local correlation tracking \citep[LCT,][]{November:1988} with a Gaussian tracking window of FWHM 1\arcsec. We first aligned the images and removed the p-mode oscillations by applying a $k-\omega$~filter with a cut-off of 5~km~s$^{-1}$.

\section{Results}
\label{results}

\begin{figure}[!b]
 \centering \includegraphics[width=\linewidth]{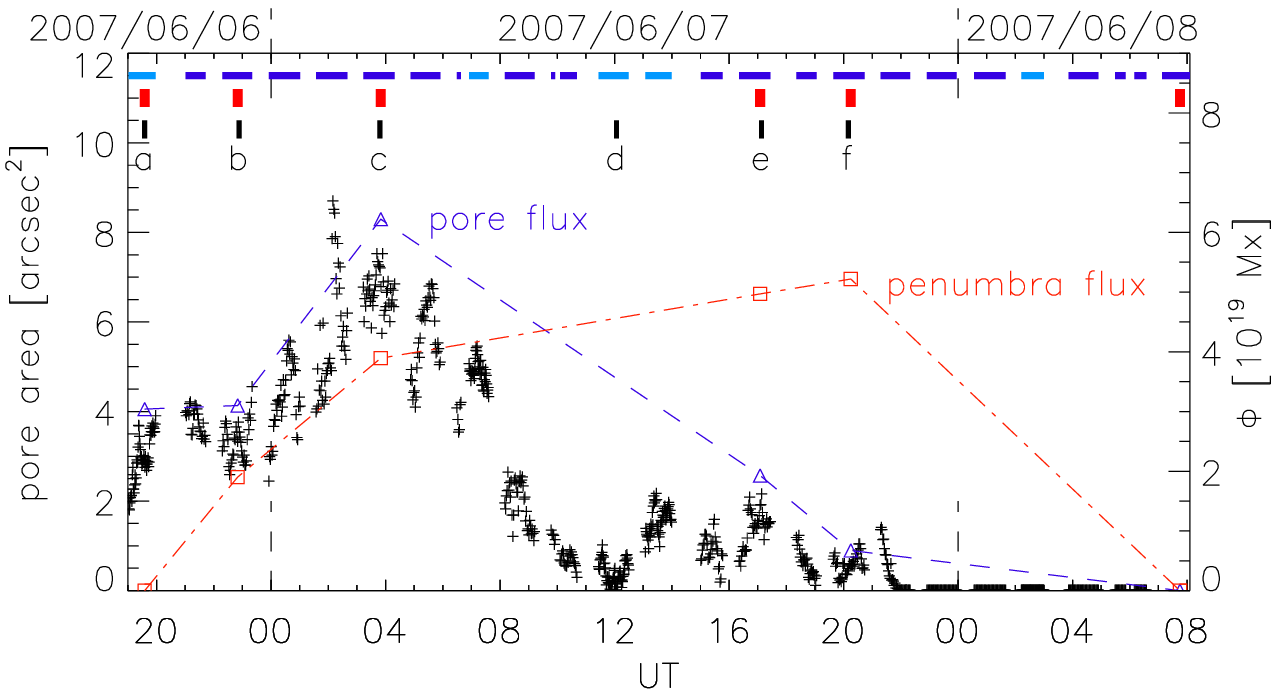}
 \caption{Evolution of the pore area with $I_\mathrm{c} < 0.5 \times I^\mathrm{QS}_\mathrm{c}$ (black $+$ symbols). Blue $\triangle$ symbols connected by the dashed line show the total positive magnetic flux ($\Phi$) in the pore. Red $\Box$ symbols connected by the dash-doted line show $\Phi$ in the penumbra. The horizontal blue lines at the top of the plot indicate the times of G-band observations, where the light blue correspond to the imaging cadence of 60~sec. The vertical red lines correspond to the times of SP scans and the vertical black lines labelled a--f show the times of images shown in Figs.~\ref{G-band},~\ref{lct}, and~\ref{sp_maps}.} 
 \label{pore_area}
\end{figure}

\begin{figure*}[!t]
 \sidecaption
 \includegraphics[width=12cm]{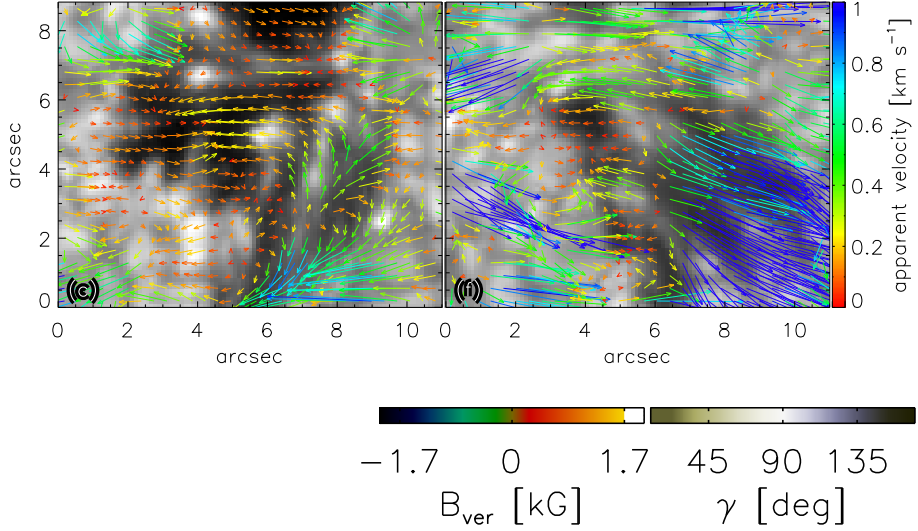}
 \caption{Arrows indicate the direction and amplitude of the apparent motions in the forming penumbra region at times around the snapshots shown in Fig.~\ref{G-band}(c, f).} 
 \label{lct} 
\end{figure*}

\begin{figure*}[!t]
 \sidecaption
 \includegraphics[width=12cm]{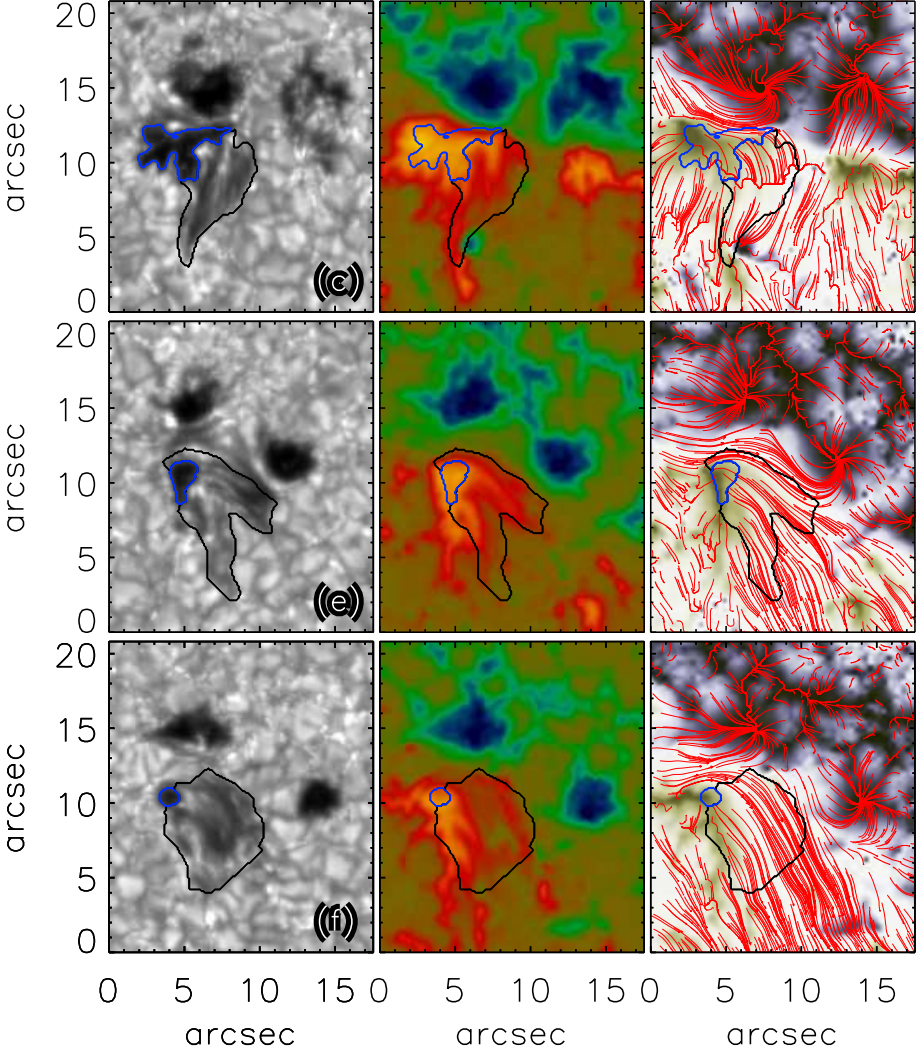}
 \caption{From left to right, we show maps of continuum intensity, $B_{\rm ver}$, and magnetic field inclination at times of the snapshots shown in Fig. 1(c, e, f). The blue and black contours indicate the areas of the studied pore and penumbra, respectively. Streamlines of the horizontal component of the magnetic field (red lines) are plotted over the magnetic field inclination maps.}
 \label{sp_maps}
\end{figure*}

In Fig.~\ref{G-band}, we show the pore and penumbra evolution in AR~10960 at selected stages. A video attached to Fig.~\ref{G-band} is available in the online edition. At the beginning of the studied period (a), the pore with a small light bridge and no penumbra is observed. Approximately two hours later (b), a penumbra segment west from the pore can be observed; the penumbral filaments, which are not directly adjacent to the pore, have south-north (S-N, cf. Fig.~\ref{G-band}f) orientation (heads of the filaments are located on the southern end). The animation on the evolution shows that this
penumbral segment in not coupled to the investigated pore. Later (c), new penumbral filaments appear with their heads - penumbral grains - connected to the pore boundary. Initially, these filaments have a N-S orientation (c), and at later stages (d, e, f), they show an E-W orientation.

The evolution of the pore area (pixels with $I_\mathrm{c} < 0.5 \times I^\mathrm{QS}_\mathrm{c}$) shown in Fig.~\ref{G-band} is shown in Fig.~\ref{pore_area}. The pore area increases until 4:00~UT on 7 June (Fig.~\ref{G-band}c). Afterwards, the pore area decreases. The evolution of the pore size is closely related to the evolution of penumbral filaments that are directly adjacent to the pore. These filaments appear around 2:00~UT on 7 June (see the animation), and we immediately observe proper motions of penumbral bright grains into the pore \citep[as observed in stable and forming penumbrae,][]{Wang:1992, Sobotka:1999, Marquez:2006, Jurcak:2015}. In Fig.~\ref{lct}, we show the proper motions in the close surrounding of the studied pore at the beginning of the penumbra formation (left) and towards the end of the pore lifetime (right). A characteristic property of a penumbra is clearly depicted: the apparent inward (outward) motion of the inner (outer) penumbra. 

In Fig.~\ref{sp_maps}, we show maps of the magnetic field configuration in the area of the forming penumbra for the three SP scans that captured the evolving pore-penumbra system. The first SP scan of the region where the penumbra is attached to the pore is shown in the upper row (c). The penumbral filaments are directed away from the polarity inversion line and the field lines (shown in the inclination map) do not connect the pore-penumbra system with the pores of opposite polarity located north and west of it. Fig. 4, middle row (e), shows the magnetic field configuration around 17:10 UT. At that time, a transient second penumbral segment develops oriented along the polarity inversion line, partially connected by the field lines to the west pore of opposite polarity around (12\arcsec, 12\arcsec). Yet, three hours later (bottom row of Fig. 4, f), the penumbral filaments are
no longer oriented westwards and the magnetic field creating the penumbra is not longer connected with the west pore of opposite polarity.

At all times, the pore-penumbra system under study belongs to the same polarity in this bipolar region. The positive magnetic flux in the penumbra increases in time (Fig. 2), while the magnetic flux in the pore diminishes. This indicates that the penumbra grows at the expense of the pore magnetic flux as signatures of emerging flux are not seen during the relevant time span. The maximum flux found in the penumbra reaches 84\% of the maximum flux in the pore, that is, even if some magnetic flux of the pore was cancelled out by the nearby regions of opposite polarity, the majority is transformed into the penumbral magnetic flux. Therefore, the fundamental cause of the pore disappearance is its transformation into the penumbra.

The process of pore disappearance is not straightforward. At some stages there are almost no regions with 
$I_\mathrm{c} < 0.5 \times I^\mathrm{QS}_\mathrm{c}$ (Fig.~\ref{G-band}d), while at later times these regions revive again (Fig.~\ref{G-band}e). During the whole sequence, we observe proper motions of penumbral bright grains into the pore (Fig.~\ref{lct}). As shown in \citet{Jurcak:2015}, such motions `displace' the UP boundary towards the umbral core in the case of a forming sunspot penumbra, until the B$_{ver}$ reaches the B$_{ver}^{stable}$ value. Analogously, the area of the studied pore decreases at the expense of the evolving penumbra. Since the strongest vertical component of the magnetic field reaches only 1.4~kG in the studied pore, the stable boundary between the pore and the forming penumbra cannot establish and the penumbral magneto-convective mode takes over in the pore area. 

The pore disappearance at the expense of the evolving penumbra can be also understood as a mechanism for an orphan penumbra formation. The predominantly vertical magnetic field of the pore becomes horizontal and we observe an orphan penumbra for a certain time before its disappearance. These horizontal fields creating the orphan penumbra submerge in time, possibly due to downward pumping as described by \citet{Tobias:1998}. This is an alternative to the formation of an orphan penumbra by a flux emergence that is blocked by an overlaying magnetic field \citep{Zuccarello:2014} or by the relation of an orphan penumbra to an active region filament \citep{Kuckein:2012, Buehler:2016}.

\section{Conclusions}
\label{discussion}

We describe the evolution of a penumbra at the boundary of a small pore. The penumbra eventually colonises the pore area leading to its extinction. We link these observations with previous studies that found constant values of the vertical component of magnetic field strength on UP boundaries of stable sunspots \citep{Jurcak:2011} and how this canonical $B_{\rm ver}^{\rm stable}$ value establishes a demarcation line in the forming UP boundary \citep{Jurcak:2015}. 

The studied pore has the vertical component of the magnetic field around 1.4~kG at its maximum (Fig.~\ref{sp_maps}) which is smaller than the $B_{\rm ver}^{\rm stable}$ of about 1.8~kG \citep{Jurcak:2011, Jurcak:2015}. Therefore, the protrusion of penumbral magneto-convection into the pore area is not hindered, a stable pore-penumbra boundary does not establish, and the pore is eventually colonised by the penumbra. As the penumbral bright grains protrude into the pore, the magnetic flux of the pore is transformed into penumbral magnetic flux, and the predominantly vertical field becomes horizontal. This scenario describes a mechanism by which an orphan penumbra forms.

\begin{acknowledgements}

The support from GA~CR~14-04338S and RVO:67985815 is gratefully acknowledged. N.B.G. acknowledges financial support by the Senatsausschuss of the Leibniz-Gemeinschaft, Ref.-No. SAW-2012-KIS-5. R.R. acknowledges financial support by the DFG grant RE 3282/1-1 and by the Spanish Ministry of Economy and Competitiveness through project AYA2014-60476-P. Hinode is a Japanese mission developed and launched by ISAS/JAXA, with NAOJ as domestic partner and NASA and STFC (UK) as international partners. It is operated by these agencies in cooperation with ESA and NSC (Norway).

\end{acknowledgements}

\bibliographystyle{aa}
\bibliography{manuscript}

\begin{thebibliography}{24}
\expandafter\ifx\csname natexlab\endcsname\relax\def\natexlab#1{#1}\fi

\bibitem[{{Buehler} {et~al.}(2016){Buehler}, {Lagg}, {van Noort}, \&
  {Solanki}}]{Buehler:2016}
{Buehler}, D., {Lagg}, A., {van Noort}, M., \& {Solanki}, S.~K. 2016, \aap,
  589, A31

\bibitem[{{Jur{\v c}{\'a}k}(2011)}]{Jurcak:2011}
{Jur{\v c}{\'a}k}, J. 2011, \aap, 531, A118

\bibitem[{{Jur{\v c}{\'a}k} {et~al.}(2014{\natexlab{a}}){Jur{\v c}{\'a}k},
  {Bello Gonz{\'a}lez}, {Schlichenmaier}, \& {Rezaei}}]{jurcak:2014a}
{Jur{\v c}{\'a}k}, J., {Bello Gonz{\'a}lez}, N., {Schlichenmaier}, R., \&
  {Rezaei}, R. 2014{\natexlab{a}}, \pasj

\bibitem[{{Jur{\v c}{\'a}k} {et~al.}(2015){Jur{\v c}{\'a}k}, {Bello
  Gonz{\'a}lez}, {Schlichenmaier}, \& {Rezaei}}]{Jurcak:2015}
{Jur{\v c}{\'a}k}, J., {Bello Gonz{\'a}lez}, N., {Schlichenmaier}, R., \&
  {Rezaei}, R. 2015, \aap, 580, L1

\bibitem[{{Jur{\v c}{\'a}k} {et~al.}(2014{\natexlab{b}}){Jur{\v c}{\'a}k},
  {Bellot Rubio}, \& {Sobotka}}]{jurcak:2014}
{Jur{\v c}{\'a}k}, J., {Bellot Rubio}, L.~R., \& {Sobotka}, M.
  2014{\natexlab{b}}, \aap, 564, A91

\bibitem[{{Kosugi} {et~al.}(2007){Kosugi}, {Matsuzaki}, {Sakao}, {Shimizu},
  {Sone}, {Tachikawa}, {Hashimoto}, {Minesugi}, {Ohnishi}, {Yamada}, {Tsuneta},
  {Hara}, {Ichimoto}, {Suematsu}, {Shimojo}, {Watanabe}, {Shimada}, {Davis},
  {Hill}, {Owens}, {Title}, {Culhane}, {Harra}, {Doschek}, \&
  {Golub}}]{Kosugi:2007}
{Kosugi}, T., {Matsuzaki}, K., {Sakao}, T., {et~al.} 2007, \solphys, 243, 3

\bibitem[{{Kuckein} {et~al.}(2012){Kuckein}, {Mart{\'{\i}}nez Pillet}, \&
  {Centeno}}]{Kuckein:2012}
{Kuckein}, C., {Mart{\'{\i}}nez Pillet}, V., \& {Centeno}, R. 2012, \aap, 539,
  A131

\bibitem[{{Leka} {et~al.}(2009){Leka}, {Barnes}, \& {Crouch}}]{Leka:2009}
{Leka}, K.~D., {Barnes}, G., \& {Crouch}, A. 2009, in Astronomical Society of
  the Pacific Conference Series, Vol. 415, The Second Hinode Science Meeting:
  Beyond Discovery-Toward Understanding, ed. B.~{Lites}, M.~{Cheung},
  T.~{Magara}, J.~{Mariska}, \& K.~{Reeves}, 365

\bibitem[{{Leka} \& {Skumanich}(1998)}]{Leka:1998}
{Leka}, K.~D. \& {Skumanich}, A. 1998, \apj, 507, 454

\bibitem[{{Lim} {et~al.}(2013){Lim}, {Yurchyshyn}, {Goode}, \&
  {Cho}}]{Lim:2013}
{Lim}, E.-K., {Yurchyshyn}, V., {Goode}, P., \& {Cho}, K.-S. 2013, \apjl, 769,
  L18

\bibitem[{{Lites} {et~al.}(1995){Lites}, {Low}, {Mart\'{\i}nez Pillet},
  {Seagraves}, {Skumanich}, {Frank}, {Shine}, \& {Tsuneta}}]{Lites:1995}
{Lites}, B.~W., {Low}, B.~C., {Mart\'{\i}nez Pillet}, V., {et~al.} 1995, ApJ,
  446, 877

\bibitem[{{M{\'a}rquez} {et~al.}(2006){M{\'a}rquez}, {S{\'a}nchez Almeida}, \&
  {Bonet}}]{Marquez:2006}
{M{\'a}rquez}, I., {S{\'a}nchez Almeida}, J., \& {Bonet}, J.~A. 2006, \apj,
  638, 553

\bibitem[{{November} \& {Simon}(1988)}]{November:1988}
{November}, L.~J. \& {Simon}, G.~W. 1988, \apj, 333, 427

\bibitem[{{Rempel} {et~al.}(2009){Rempel}, {Sch{\"u}ssler}, {Cameron}, \&
  {Kn{\"o}lker}}]{Rempel:2009b}
{Rempel}, M., {Sch{\"u}ssler}, M., {Cameron}, R.~H., \& {Kn{\"o}lker}, M. 2009,
  Science, 325, 171

\bibitem[{{Rezaei} {et~al.}(2012){Rezaei}, {Bello Gonz{\'a}lez}, \&
  {Schlichenmaier}}]{Rezaei:2012}
{Rezaei}, R., {Bello Gonz{\'a}lez}, N., \& {Schlichenmaier}, R. 2012, \aap,
  537, A19

\bibitem[{{Rucklidge} {et~al.}(1995){Rucklidge}, {Schmidt}, \&
  {Weiss}}]{Rucklidge:1995}
{Rucklidge}, A.~M., {Schmidt}, H.~U., \& {Weiss}, N.~O. 1995, \mnras, 273, 491

\bibitem[{{Ruiz Cobo} \& {del Toro Iniesta}(1992)}]{Cobo:1992}
{Ruiz Cobo}, B. \& {del Toro Iniesta}, J.~C. 1992, ApJ, 398, 375

\bibitem[{{Sobotka} {et~al.}(1999){Sobotka}, {Brandt}, \&
  {Simon}}]{Sobotka:1999}
{Sobotka}, M., {Brandt}, P.~N., \& {Simon}, G.~W. 1999, A\&A, 348, 621

\bibitem[{{Tobias} {et~al.}(1998){Tobias}, {Brummell}, {Clune}, \&
  {Toomre}}]{Tobias:1998}
{Tobias}, S.~M., {Brummell}, N.~H., {Clune}, T.~L., \& {Toomre}, J. 1998,
  \apjl, 502, L177

\bibitem[{{Tsuneta} {et~al.}(2008){Tsuneta}, {Ichimoto}, {Katsukawa}, {Nagata},
  {Otsubo}, {Shimizu}, {Suematsu}, {Nakagiri}, {Noguchi}, {Tarbell}, {Title},
  {Shine}, {Rosenberg}, {Hoffmann}, {Jurcevich}, {Kushner}, {Levay}, {Lites},
  {Elmore}, {Matsushita}, {Kawaguchi}, {Saito}, {Mikami}, {Hill}, \&
  {Owens}}]{Tsuneta:2008}
{Tsuneta}, S., {Ichimoto}, K., {Katsukawa}, Y., {et~al.} 2008, \solphys, 249,
  167

\bibitem[{{Wang} \& {Zirin}(1992)}]{Wang:1992}
{Wang}, H. \& {Zirin}, H. 1992, \solphys, 140, 41

\bibitem[{{Zirin} \& {Wang}(1991)}]{Zirin:1991}
{Zirin}, H. \& {Wang}, H. 1991, Advances in Space Research, 11, 225

\bibitem[{{Zuccarello} {et~al.}(2014){Zuccarello}, {Guglielmino}, \&
  {Romano}}]{Zuccarello:2014}
{Zuccarello}, F., {Guglielmino}, S.~L., \& {Romano}, P. 2014, \apj, 787, 57

\bibitem[{{Zwaan}(1987)}]{Zwaan:1987}
{Zwaan}, C. 1987, \araa, 25, 83

\end{thebibliography}

\end{document}